\documentclass[11pt]{article}
\usepackage[pagewise]{lineno}
\usepackage{amsmath,amssymb,amsthm}
\usepackage[all]{xy}
\usepackage{authblk}
\usepackage[utf8]{inputenc}
\usepackage{fontenc}
\usepackage{mathtools, nccmath, etoolbox}
\usepackage[a4paper]{geometry}
\usepackage[hidelinks]{hyperref} 
\usepackage{graphicx}
\usepackage[sort]{natbib} 
\usepackage{mathdots}
\usepackage{xcolor}
\usepackage{comment}
\usepackage{tikz}
\usetikzlibrary{matrix}
\theoremstyle{plain}
\newtheorem{theorem}{Theorem}
\newtheorem{proposition}[theorem]{Proposition}

\theoremstyle{definition}
\newtheorem{definition}[theorem]{Definition}
\newtheorem{example}{Example}
\theoremstyle{remark}
\newtheorem*{remark*}{Remark}

\def\r{\ensuremath{\mathbb{R}}}
\def\rk{{\mathbb R}^{k}}
\def\tkq{T^1_kQ}

\def\rkq{\rk \times Q}
\def\rkq{\rk \times Q}

\def\derpar#1#2{\displaystyle\frac{\partial{#1}}{\partial{#2}}}

\providecommand{\keywords}[1]
{
  \small	
  \textbf{\textit{Keywords---}} #1
}

\begin{document}


\title{ Jacobi equation for field theories and a geometric variational description of dissipation}

\author[1]{David Mart{\'\i}n de Diego}
\author[2]{Najma Mosadegh}
\affil[1]{\protect\raggedright Instituto de Ciencias Matem\'aticas (CSIC-UAM-UC3M-UCM), Campus
de Cantoblanco, UAM Nicolas Cabrera, 1528049 Madrid, Spain\thanks{david.martin@icmat.es}}
\affil[2]{\protect\raggedright
Department of Mathematic, Azarbaijan Shahid Madani University, Tabriz, Iran\thanks{n.mosadegh@azaruniv.ac.ir}}
	
	\date{\today}

\maketitle

\abstract{In this paper we give a geometric description of the Jacobi equations associated to a first-order Lagrangian field theory using a  prolongation of the Lagrangian $L$ on a $k$-cosymplectic formulation. Moreover, using an appropriate modification of the prolonged Lagrangian, we obtain a variational formulation of field theories with dissipation.}


\maketitle

\section{Introduction}

The field equations are obtained in a rather intuitive way by making use of
the calculus of variations, and that is precisely the approach developed by T. de Donder
who extended the Hamiltonian formulation for mechanics due to E. Cartan.
This theory was discussed later by H. Weyl \cite{weyl} so that the theory was known as
the De Donder-Weyl theory. The introduction of the notions of fiber bundles
and connections by C. Ehresmann \cite{ehres} provided the additional tool for developing
the geometrical arena for a further step in the study of classical field theories.
Lagrangian field theories are usually framed in the context of jet bundles.
These spaces are fiber bundles over a base manifold where each fiber can be
understood as keeping information of the configuration of a field and its derivative. 
More precisely, let $\pi : Y \rightarrow X $ be a fiber bundle, the configuration bundle,
with local adapted coordinates $(x^{\mu}, q^i)$, i.e. $\pi(x^{\mu}, q^i) = (x^{\mu})$. A field is understood as a local section of this bundle, $\sigma : U \subset X\rightarrow Y, \pi \circ \sigma=  {\rm Id}_U$ . 
The field equations are determined given a first-order Lagrangian, $L : J^1\pi \rightarrow {\mathbb R}$ where $J^1\pi$, is the bundle of all equivalence classes of sections whose derivatives agree up to order $1$ \cite{saunders}, in coordinates $ L (x^{\mu}, q^i, q^i_{\mu})$. Then, one defines the action functional
\begin{eqnarray}
S[\sigma] = \int_U  L( j^1 \sigma)\; \eta,  
\end{eqnarray}
where $j^1 \sigma$ is the 1-jet prolongation of the section $\sigma$, and
$\eta$ is a volume form on $X$.  Locally 
$j^1\sigma (x)= (x^{\mu}, q^i(x), \frac{\partial q^i}{\partial x^{\mu}}(x))$.
Taking variations of this action leads to the Euler-Lagrange field equations
\begin{eqnarray}\label{eq:field}
\sum_{\mu}\frac{d}{d x^{\mu}}\left(\frac{\partial L}{\partial q^i_{\mu}}\right) - \frac{\partial L}{\partial q^i}=0,  
\end{eqnarray}
A crucial role for the study of variational problems is played by Jacobi equations. For instance in classical Riemannian geometry \cite{doCarmo}, when we consider 1-parameter families of Riemannian geodesics, the Jacobi fields correspond to the velocity fields of transversal curves along one fixed geodesic and are characterized by the Jacobi equation.

The Jacobi equation can be naturally generalized to Lagrangian systems by taking variations of the Euler-Lagrange equations. Something similar happens for field theories. In this paper,   we will give a geometric interpretation of the Jacobi equation lifting the Lagrangian $L$ obtaining now a new Lagrangian $\tilde{L}(x^{\mu}, q^i, v^i, q^i_{\mu}, {v}^i_{\mu})$ with corresponding Euler-Lagrange field equations: 
\begin{eqnarray}\label{eq: extendedfield}
\sum_{\mu}\frac{d}{d x^{\mu}}\left(\frac{\partial \tilde{L}}{\partial q^i_{\mu}}\right) - \frac{\partial \tilde{L}}{\partial q^i}=0,  \qquad 
\sum_{\mu}\frac{d}{d x^{\mu}}\left(\frac{\partial \tilde{L}}{\partial {v}^i_{\mu}}\right) - \frac{\partial \tilde{L}}{\partial {v}^i}=0,
\end{eqnarray}
where the second equation corresponds to Equations (\ref{eq:field}) and the first one is precisely the Jacobi field equations (see also \cite{CFT}). The Euler-Lagrange equations are intrinsically derived using the $k$-cosymplectic formalism \cite{MR3410116,dne,MR2915598,MR2794297}.

Moreover, as an interesting consequence of this geometric construction it is related with the variational description of variational problems adding extra-variables. In the paper \cite{bateman} H. Bateman discusses when a dissipative
system can be described in a variational way:
\begin{quote}
``A given set of differential equations is always included in a set derivable from a variational principle. In the case of a set of equations
representing a dissipative physical system the complementary set of
equations may represent a second physical system which absorbs the
energy dissipated by the first. This is illustrated by an example in
which the total kinetic energy is never negative only when the initial
conditions for the second system are related to those for the first". 
\end{quote}
This approach was successfully used in the modelization of dissipative forces from a pure variational perspective (see \cite{Galley1} and references therein). 
In our case, we will derive a similar technique modifying the Lagrangian function $\tilde{L}$ and  deriving a variational description of any field equations admitting extra-terms determined by functions $F_i$ and $F^{\mu}_i$ which describe, for instance, dissipative behavior. In other words, we will give a purely variational description of systems given by field equations of the type
\begin{eqnarray}\label{eq:fieldforced}
\sum_{\mu}\frac{d}{d x^{\mu}}\left(\frac{\partial L}{\partial q^i_{\mu}}-{F}^{\mu}_i\right) - \frac{\partial L}{\partial q^i}=F_i,  
\end{eqnarray}
where $F_i$ and $F^{\mu}_i$ depend on $(x^{\mu}, q^i, q^i_{\mu})$.
The dissipative field theories is a topic recently studied in geometric terms in  \cite{diss1,diss2,diss3,diss4} among others. Our approach, since it is variational from construction, allows us to  directly apply results derived for the cases of Lagrangian system to systems with dissispation (for instance, see recent results about variational error analysis for forced systems in \cite{sato1}).  

\section{Geometric preliminaries}\label{gepre}

In this paper we will use for simplicity   the $k$-cosymplectic approach to classical field theories (see \cite{dne,MR3410116} and references therein).

\subsection{The manifold $\rk \times T^1_kQ$}

 Let $Q$ be an $n$-dimensional manifold and $\tau_Q\colon TQ\to Q$ the canonical tangent bundle projection given by $\tau_Q(v_q)=q$ where $v_q\in T_qQ$. $TQ$ is the space to describe  dynamics in classical mechanics, but for field theories we need to define   $\tkq$, the Whitney sum $TQ
\oplus\stackrel{k}{\ldots}\oplus TQ$ of $k$ copies of $TQ$ to take into account partial derivatives of the field variable $q$ with respect $k$-independent variables. Denote by   $\tau^k_Q\colon\tkq\to Q$, the projection defined by $\tau^k_Q({v_1}_q,\ldots, {v_k}_q)=q$, where ${v_\mu}_q\in T_qQ,\, \mu=1,\ldots, k$.
$\tkq$ is usually called the tangent bundle of $k^1$-velocities of $Q$, since for any map
$\sigma\colon\rk\to Q$   with the source at $\mathbf{0}\in\rk$, we have the following identification 
$$\begin{array}{ccc}
J_\mathbf{0}^1(\rk,Q) & \equiv & T^1_kQ= TQ\,
\oplus\stackrel{k}{\ldots}\oplus\, TQ\\\noalign{\medskip}
j^1_{\mathbf{0},q}\sigma &\equiv & ({v_1}_q,\ldots, {v_k}_q)
\end{array}$$
 where $q=\sigma(\mathbf{0})$  and ${v_\mu}_q=T_{\mathbf{0}}\sigma\left(\frac{\partial}{\partial x^\mu}\Big|_{\mathbf{x}=\mathbf{0}}\right)=\frac{\partial \sigma}{\partial x^\mu}(\mathbf{0})$ and  ${\bf x}=(x^1,\ldots, x^k)$ being the standard coordinates on $\rk$ (see \cite{morimoto}).

For more general field theories we need to introduce the jet manifold $J^1\hat{\pi}^k_Q$ of 1-jets of sections of the trivial
bundle $\hat{\pi}^k_Q:\rk \times Q \to \rk$. This space  is diffeomorphic to $\rk
\times  T^1_kQ$, via the diffeomorphism given by
$$
\begin{array}{rcl}
J^1\hat{\pi}^k_Q & \longrightarrow  & \rk   \times T^1_kQ \\
\noalign{\medskip} j^1_{\bf x}\phi= j^1_{\bf x}(\hbox{id}_{\rk}\times\phi_Q) & \longmapsto & (
{\bf x},v_1, \ldots ,v_k)
\end{array}
$$
where $\phi_Q: \rk \stackrel{\phi}{\longrightarrow}  \rkq \stackrel{\hbox{\tiny proj}_2}{\longrightarrow}Q
$, and
$$
v_\mu=T_{\mathbf{x}}\phi_Q\left(\displaystyle\frac{\partial}{\partial
x^\mu}\Big\vert_\mathbf{x}\right)\in T_{\phi_Q({\bf x})}Q\, , \quad  1\leq \mu \leq k \, .
$$

Let $p_Q:\rk\times \tkq \to Q$ be the canonical projection. If $(q^i)$ are local coordinates on $U
\subseteq Q$,  then the induced local coordinates  $(x^\mu,q^i, q^i_\mu)$ on
$p_Q^{-1}(U)=\rk \times T^1_kU$ are expressed by
$$x^\mu(\mathbf{x},{v_1}_q,\ldots , {v_k}_q)   =   x^\mu; \quad
q^i(\mathbf{x},{v_1}_q,\ldots , {v_k}_q)  =q^i(q); \quad
q_\mu^i(\mathbf{x},{v_1}_q,\ldots , {v_k}_q)    =
\langle dq^i, v_{\mu_q}\rangle\, , $$ where $1\leq i \leq n,\, 1\leq \mu \leq
k$.

Throughout the paper we use the following notation for the canonical
projections
\[\xymatrix@C=13mm{\rk\times T^1_kQ\ar[r]^-{(\hat{\pi}^k_{Q})_{1,\,0}}\ar[dr]_-{(\hat{\pi}^k_{Q})_1}
& \rkq\ar[d]^-{\hat{\pi}^k_{Q}}\\
 &\r^k
}\]
 where, for $\mathbf{x}\in \rk $, $q\in Q$ and $({v_1}_q,\ldots , {v_k}_q)\in
T^1_kQ$,
$$\hat{\pi}^k_{Q}(\mathbf{x},q)=\mathbf{x}, \quad (\hat{\pi}^k_{Q})_{1,\,0}(\mathbf{x},{v_1}_q,\ldots ,
{v_k}_q)=(\mathbf{x},q), \quad (\hat{\pi}^k_{Q})_1(\mathbf{x},{v_1}_q,\ldots , {v_k}_q)=\mathbf{x}  \ .$$

\subsection{$k$-vector fields and integral sections} \label{212}

Let $M$ be an arbitrary differentiable manifold.

\begin{definition}  \label{kvector} A section ${\bf X}: M \longrightarrow T^1_kM$ of the projection
$\tau^k_M$ will be called a  {\rm $k$-vector field} on $M$.
\end{definition}

To give a $k$-vector field ${\bf X}$ is equivalent to give a
family of $k$ vector fields $X_{1}, \dots, X_{k}$. Hence in the sequel we will indistinctly write $\mathbf{X}=(X_1, \ldots, X_k)$.

\begin{definition} \label{integsect}
An {\rm integral section} of the $k$-vector field  \, $\mathbf{X}=(X_{1},
\dots, X_{k})$, passing through a point $m\in M$,  is a map
$\psi:U_\mathbf{0}\subset \r^k \rightarrow M$, defined on some neighborhood
$U_\mathbf{0}$ of $\mathbf{0}\in \rk$, such that $\psi(\mathbf{0})=m$, and
\begin{equation}\label{gintdec}
\psi_{*}(\mathbf{x})\left(\displaystyle\frac{\displaystyle\partial}{\displaystyle\partial
x^\mu}\Big\vert_\mathbf{x}\right) =T_{\bf{x}}\psi\left(\displaystyle\frac{\displaystyle\partial}{\displaystyle\partial
x^\mu}\Big\vert_\mathbf{x}\right)  =X_{\mu}(\psi (\mathbf{x})), \  \mbox{for every}
\quad \mathbf{x}\in U_\mathbf{0},\, 1\leq \mu\leq k,
\end{equation}
or,  equivalently,  $
\psi({\mathbf{0}})= {x}$ and $\psi$ satisfy
${  \mathbf{X}}\circ\psi=\psi^{(1)}$, where  $\psi^{(1)}$ is the first
prolongation of $\psi$  to $T^1_kM$, defined by
$$
\begin{array}{rccl}\label{1prolong}
\psi^{(1)}: & U_{0}\subset \r^k & \longrightarrow & T^1_kM
\\\noalign{\medskip}
 &\mathbf{x} & \longrightarrow & j^1_{\bf x}\psi\equiv \psi^{(1)}(\mathbf{x})=j^1_{\mathbf{0}}\psi_\mathbf{x}\equiv
 \left(\psi_*(\mathbf{x})\left(\derpar{}{x^1}\Big\vert_\mathbf{x}\right),\ldots,
\psi_*(\mathbf{x})\left(\derpar{}{x^k}\Big\vert_\mathbf{x}\right)\right) \, ,
 \end{array}
$$ where $\psi_\mathbf{x}({  \mathbf{y}})=\psi(\mathbf{x}+{  \mathbf{y}})$.
In coordinates, if $\psi(\mathbf{x})=(\mathbf{x}, q^i(\mathbf{x}))$ then 
\[
\psi^{(1)}(\mathbf{x})=(\mathbf{x}, q^i(\mathbf{x}), \frac{\partial q^i}{\partial x^{\mu}}(\mathbf{x}))\; , \  1\leq \mu\leq k,\ 1\leq i\leq n,
\]
where ${\mathbf x}=(    x^1, \ldots, x^k)$.

A $k$-vector field $\mathbf{X}=(X_1,\ldots , X_k)$ on $M$ is said to be {\rm
integrable} if there is an integral section passing through every
point of $M$.
\end{definition}

\subsection{Canonical structures in $\rk\times T^1_kQ$}

For a vector field $Z\in {\mathfrak X}(Q)$  we define the $k$-vertical lifts to $T^1_kQ$ by 
\[
(Z)^{V_\mu}(v_{1q}, \ldots, v_{\mu q}, \ldots, v_{kq})=
\frac{d}{ds}\Big|_{s=0} (v_{1q}, \ldots, v_{\mu q}+sZ(q), \ldots, v_{kq}), 
\]
for all $1\leq \mu\leq k$. 
Therefore, locally
\[
(Z)^{V_\mu}=Z_i\frac{\partial}{\partial q^i_\mu}
\]
where locally $Z=Z_i\frac{\partial}{\partial q^i}$.

Additionally, define the set of $k$ (1,1)-tensor fields $S^\mu$ in $T^1_k Q$ by
\[
S^\mu(w_q)(X_{w_q})= (T_{w_q}\tau_Q^k(X_{w_q}))^{V_\mu}_{w_q}
\]
for all $X_{w_q}\in T_{\omega_q}T^1_kQ$. 
$(S^1,\ldots, S^k)$ is called the canonical $k$-tangent structure of $T^1_kQ$.
Denote by $\bar{S}^\mu$ their extensions to $\rk\times T^1_kQ$. In coordinates
\[
\bar{S}^\mu=\frac{\partial}{\partial q^i_\mu}\otimes d q^i
\]

Finally, we introduce the Liouville vector field $\bar{\Delta}$ as the vector field with flow generated by dilations
\[
\begin{array}{rcl}
{\mathbb R}\times (\rk\times T^1_k Q)&\longrightarrow& \rk\times T^1_k Q\\
(s, ({\bf x}, \omega_q)&\longmapsto& ({\bf x}, e^s\omega_q)
\end{array}
\]
In local coordinates
\[
\bar{\Delta}=\sum_{i, \mu}q^i_\mu\frac{\partial}{\partial q^i_\mu}\; .
\]

Also, define the vector fields  
\[
\bar{\Delta}_\mu=
\sum_{i}q^{i}_{\mu}\frac{\partial} {\partial q^i_{\mu}}
\]
Observe that $\bar{\Delta}=\sum_{\mu}\bar{\Delta}_\mu$.

Given a $k$-vector field ${\mathbf X}= (X_1, ..., X_k)$ on  $\rk\times T^1_k Q$. If every integral curve of $X_{\alpha}$ is a prolongation $\psi^{(1)}$ of map $\psi: \rk \longrightarrow Q$ then $X= (X_1, ..., X_k)$ is called a second order partial differential equation (SOPDE for short). \\
Equivalently, a $k$-vector field ${\mathbf X}$ is a SOPDE if 
$\bar{S}^{\mu}({\mathbf X})=\bar{\Delta}_{\mu}$ and ${dx}^{\mu}(X_{\nu})=\delta^{\mu} _{\nu}$ for $1\leq \mu , \nu \leq k$. 
Locally
\[
X_\mu=\frac{\partial }{\partial x^\mu}+q^i_{\mu}\frac{\partial}{\partial q^i}+(f_{\mu})^i_{\nu}\frac{\partial}{\partial q^i_{\nu}}
\]
where $(f_{\mu})^i_{\nu}\in C^{\infty}({\mathbb R}^k\times T^1_k Q)$.

\subsection{$k$-cosymplectic structures}
To define geometrically the field equations it is necessary to introduce the  geometric structure of $k$-cosymplectic structure that extends the classical notion of cosymplectic structure for non-autonomous Lagrangian theories.  Let $M$ be a differentiable manifold of dimension $k(n+1) +n$. 
 \begin{definition}
     A $k$–cosymplectic structure is a family $(\eta_{\mu},\Omega_{\mu},V)$, $1\leq \mu\leq k$, where $\eta_{\mu}\in \Omega^1(M)$ and $ \Omega_{\mu}\in\Omega^2 (M)$, and $V$ is an $nk$-dimensional distribution on $M$ verifying that 
\begin{enumerate}
\item $\eta_1\wedge \ldots \wedge \eta_k\not= 0$, $\eta_\mu|_{V}=0$, $\Omega_\mu|_{V\times V}=0$.
\item 
 $\left(\cap_{\mu=1}^k \ker \eta_{\mu}\right) \cap
\left(\cap_{\mu=1}^k  \ker \Omega_{\mu}\right)= {0}$, and $\dim\left(\cap_{\mu=1}^k  \ker \Omega_{\mu}\right) = k.$ 
\item  All the forms $\eta_{\mu}$ and $\Omega_{\mu}$ are closed and   $V$ is integrable.
\end{enumerate}
Then $(M, \eta_{\mu}, \Omega_{\mu}, V)$  is said to be a $k$–cosymplectic manifold.
\end{definition}
Given a  $k$-cosymplectic structure $(\eta_{\mu},\Omega_{\mu},V)$, $1\leq \mu\leq k$ on $M$, then we can define  a  $k$-vector field ${\mathbf R}=(R_{1}, \ldots, R_k)$, which is called the Reeb $k$-vector field, characterized by 
\[
i_{R_{\mu}}\eta_{\nu}=\delta_{\mu\nu}\; ,\qquad  i_{R_{\mu}}\Omega_{\nu} =0 \; ,\qquad 
1\leq \mu,\nu \leq k\; .
\]

\subsection{Field equations for a Lagrangian system}

Consider the space $C^2(\rk, Q)$ of $C^2$-sections
$\sigma: \rk\rightarrow Q$. 
Given a Lagrangian function $L\in C^2(\rk\times T^1_kQ)$ we can consider the action functional 
$
{\mathcal S}_L: C^2(\rk, Q)\rightarrow {\mathbb R}$
defined by
\[
{\mathcal S}_L (\sigma)=\int_{\Omega}L (j^1_{\bf x}\sigma)d^kx
\]
where $d^k x=dx^1\wedge \ldots dx^k$ is the canonical volume form on ${\mathbb R}^k$. 

It is well known that the extremals $\sigma$ of ${\mathcal S}_L$ are characterized by:
\[
\frac{d}{ds}\Big|_{s=0}
{\mathcal S}_L (\sigma_s)=0
\]
where $\sigma_s\in C^2(\rk, Q)$ 
with $\sigma_0=\sigma$ and $s\in (-\epsilon, \epsilon)$ with $\epsilon>0$.
It is well known that these critical sections are the solutions  of the Euler-Lagrange field equations: 
 \begin{eqnarray*}
\sum_{\mu}\frac{d}{d x^{\mu}}\left(\frac{\partial L}{\partial q^i_{\mu}}\right) - \frac{\partial L}{\partial q^i}=0. 
\end{eqnarray*}

The Lagrangian $L$ is said to be regular if the matrix 
\[
\left( \frac{\partial^2 L}{\partial q^i_\mu\partial q^j_{\nu}}\right)
\]
is  non-singular  at every point of ${\mathbb R}^k\times T^1_k Q$.

For our purposes, it would be necessary to introduce an intrinsic version of the Euler-Lagrange field equations using the $k$-cosymplectic formalism. 

To this end, now we consider that a family of forms $\Theta^\mu_L\in \Omega^1({\mathbb R}^k\times T^1 _k Q)$ , $1\leq \mu \leq k$, is defined by using the canonical structure previously defined as follows
\begin{eqnarray}
\Theta^\mu _L={d}L\circ \overline{S}^\mu,
\end{eqnarray}
so, we introduce the 2-forms $\Omega^\mu _L=-{d}\Theta^\mu _L$. Then in the induced coordinates
\begin{eqnarray}
\Theta^\mu _L=\frac{\partial L}{\partial q^i _\mu}d{q}^i,\ \ \ \ \Omega^\mu _L={d}q^i\wedge {d}\left(\frac{\partial L}{\partial q^i _\mu}\right)=\frac{\partial^2 L}{\partial q^j \partial q^i _\mu}{d}q^i\wedge {d}q^j+ \frac{\partial^2 L}{\partial q^j _\gamma \partial q^i _\mu}{d}q^i\wedge {d}q^j_\gamma
\end{eqnarray}
Also, we  recall the Energy Lagrangian function associated to $L$ as $E_L=\bar{\Delta} (L)-L$. In local coordinates
\begin{eqnarray}
E_L=q_i^\mu\frac{\partial L}{\partial q^i _\mu}-L.
\end{eqnarray}
From the above geometrical structure we recall the following definition that is introduced in \cite{dne}. 
Define
\begin{eqnarray}
V={\ker}((\pi_{\rk})_{1,0})_{\ast} = \textsf{span}\{\frac{\partial }{\partial v^i}, ..., \frac{\partial }{\partial v^k}\}
\end{eqnarray}
 the vertical distribution of the vector bundle $(\pi_{\rk})_{1,0}$. Then, given a regular Lagrangian function $L\in C^{\infty}({\mathbb R}^k \times T^1 _k Q)$  then  $(dx^{\mu},\Omega_L^1, ..., \Omega^k_L, V)$ is a $k$-cosymplectic structure on ${\mathbb R}^k \times T^1 _k Q$.

 With the above geometric elements the geometric $k$-cosymplectic description of the Euler-Lagrange field equations of $L\in C^{\infty}({\mathbb R}^k \times T^1 _k Q)$ is as follows:
\begin{equation}
      \begin{array}{cc}
       dx^{\mu}(X_{\nu})=\delta^{\mu} _{\nu}    &  1\leq \mu, \nu \leq k\\ \\
       \sum_{\mu=1}^k i_{X_\mu} \Omega^{\mu} _L=dE_L +\sum \frac{\partial L}{\partial x^{\mu}} dx^{\mu}\\    & 
      \end{array}
\end{equation}
as a geometric version of the Euler-Lagrange field equations in terms of the $k$-cosymplectic structure where a set of $k$-vector field ${\mathbf X}=(X^1, ..., X^k)$ denotes the solution of it.
  If $L$ is regular, then ${\mathbf X}$ is a SOPDE and  if it is integrable, its integral sections  are solutions to the Euler-Lagrange equations for $L$ (see \cite{dne} and references therein).

\subsection{The canonical isomorphism between $TT^1_kQ$ and $T^1_kTQ$  }
The double tangent bundle $TTQ$ admits two vector bundle structures \cite{Yano-Ishihara,deLeon-Rodrigues}.
The first is the canonical one given by the vector bundle projection $\tau_{TQ}:TTQ\rightarrow TQ$. For the second vector bundle structure, the vector bundle projection is just the tangent map to $\tau_{Q}$, that is, $T\tau_{Q}:TTQ\rightarrow TQ$ and, the last case the addition operation on the fibers is just the tangent map $T(+):TTQ\times_{TQ} TTQ\rightarrow TTQ$ of the addition operation $(+):TQ\times_{Q}TQ\rightarrow TQ$ on the fibers of $\tau_{Q}$.

The canonical involution $\kappa_{Q}:TTQ\rightarrow TTQ$ is a vector bundle isomorphism (over the identity of $TQ$) between the two previous vector bundles. In fact, $\kappa_{Q}$ is characterized by the following condition: let $\Phi:U\subseteq \r^{2}\rightarrow Q$ be a smooth map, with $U$ an open subset of ${\mathbb R}^{2}$
\begin{equation*}
(t,s)\mapsto \Phi(t,s)\in Q.
\end{equation*}
Then,
\begin{equation}\label{def:can:invol}
\kappa_{Q}\left( \frac{d}{dt}\frac{d}{ds} \Phi(t,s) \right)= \frac{d}{ds}\frac{d}{dt} \Phi(t,s).
\end{equation}
So, we have that $\kappa_{Q}$ is an involution of $TTQ$, that is, $\kappa_{Q}^{2}=id_{TTQ}$.

In fact, if $(q^{i},\dot{q}^{i})$ are canonical fibered coordinates on $TQ$ and $(q^{i},\dot{q}^{i},v^{i},\dot{v}^{i})$ are the corresponding local fibered coordinates on $TTQ$ then
\begin{equation}\label{def:local:can:invol}
\kappa_{Q}(q^{i},\dot{q}^{i},v^{i},\dot{v}^{i})=(q^{i},v^{i},\dot{q}^{i},\dot{v}^{i}).
\end{equation}

It is easy to extend the canonical involution to $ T^1_kQ$ defining the map
$\kappa^k_{Q}: T^1_kTQ\rightarrow TT^1_kQ  $ as follows.  
Let $\Phi:\rk\times {\mathbb R}\rightarrow Q$ be a smooth map
\begin{equation*}
({\mathbf t},s)\mapsto \Phi({\mathbf t},s)\in Q.
\end{equation*}
We denote by $\Phi_s({\mathbf t})=\Phi_{\mathbf t}(s)=\Phi({\mathbf t},s)$.
Then, we define 
\begin{equation}\label{def:can:invol}
\kappa^k_{Q}\left(\left(\frac{d\Phi_{\mathbf t}}{ds}(s)\right)^{(1)} ({\mathbf t})\right) =    \frac{d \Phi_s^{(1)}({\mathbf t})}{ds}(s)  \; .
\end{equation}
Observe that 
\[
 \frac{d \Phi_s^{(1)}({\mathbf t})}{ds}: \r\rightarrow TT^1_kQ\; ,\qquad \frac{d\Phi_{\mathbf t}}{ds}(s): \rk\rightarrow TQ,\quad \left(\frac{d\Phi_{\mathbf t}}{ds}(s)\right)^{(1)} ({\mathbf t})\in T^1_kTQ
\]
In local coordinates
\[
\kappa^k_{Q}(q^i, {v}^i;  q^i_A , {v}^i_A)=(q^i, q^i_A; {v}^i, {v}^i_A)\, .
\]

$\kappa^k_{Q}$ may be also characterized in a more intrinsic way, using the theory of complete and vertical lifts to $TQ$ and $T^1_kQ$.
Given a  {\rm $k$-vector field} on $Q$, 
$\mathbf{X}=(X_1, \ldots, X_k)$, we can consider the 
{\rm $k$-vector fields} $\mathbf{X}^c$ and  $\mathbf{X}^v$ on $TQ$ defined using complete and vertical lifts, that is,  
\[
\mathbf{X}^c=(X^c_1, \ldots, X^c_k), \qquad \mathbf{X}^v=(X^v_1, \ldots, X^v_k)\; .
\]

Indeed, if $X$ is a {\rm $k$-vector field} on $Q$
\[
\kappa^k_{Q}\circ \mathbf{X}^c=T{\mathbf X}, \quad \kappa^k_{Q}\circ \mathbf{X}^v=\widetilde{\mathbf X}^{v},
\]
where $T{\mathbf X}:TQ\rightarrow TT^1_kQ$ is the tangent map to $X$ (a section of the vector bundle $T\tau^k_{Q}$) and $\widetilde{\mathbf X}^{v}:TQ\rightarrow TT^1_kQ$ is the section of the vector bundle $T\tau^k _{Q}$ given by
\begin{equation*}
	\widetilde{\mathbf X}^{v}(u)=\left((T_{q}0)(u)+X_1^{v}(0(q)),\ldots, (T_{q}0)(u)+X_k^{v}(0(q))\right) \quad u\in T_{q}Q,
\end{equation*}
with $0:Q\rightarrow TQ$ the zero section.

\section{The prolongation of the Lagrangian $L$ to  the $k$-cosymplectic manifold $\rk \times T^1_k TQ$} 
In this section, we will derive the Jacobi field equations as the Euler-Lagrange field equations corresponding to a Lagrangian function  $\widetilde{L}$  defined on ${\mathbb R}^k\times T^1 _k TQ$.

Given a regular Lagrangian $L: {\mathbb R}^k\times T^1_k Q\rightarrow {\mathbb R}$ consider its complete lift $L^C: {\mathbb R}^k\times TT^1_k Q\rightarrow {\mathbb R}$ defined by
\[
L^C({\mathbf x}, \omega_q, V_{ \omega_q})=\langle dL_{\mathbf x}(  \omega_q), V_{ \omega_q}\rangle\; , 
\]
for all $\omega_q\in T^k_1 Q$ and $V_{ \omega_q}\in T_{ \omega_q}T^1_kQ$ and where 
$L_{\mathbf x}(\omega_q)=L({\mathbf x}, \omega_q)$.

Finally, the lifted Lagrangian $\tilde{L}: {\mathbb R}^k\times {T^1 _{k}TQ}\longrightarrow {\mathbb R}$ is defined by 
 $$\widetilde{L}:= L^C\circ \left(\hbox{id}_{\rk}\times\kappa^k_Q\right)\; .$$ Then, in a coordinate system $(x^{\mu}, q^i, q^i_{\mu})$ in $\rk \times T^1 _k Q$, we have

\begin{equation}
\widetilde{L}\big(x^{\mu}, q^i, {v}^{i}, q^i_{\mu}, {v}^{i}_{\mu}\big)={L}^C\big(x^{\mu}, q^i, q^i_{\mu},{v}^{i}, {v}^{i}_{\mu}\big).
\label{eq:1}
\end{equation}
Locally
\begin{equation*}
\widetilde{L}\big(x^{\mu}, q^i, {v}^{i}, q^i_{\mu}, {v}^{i}_{\mu}\big)=
\frac{\partial L}{\partial q^{i}}\big(x^{\mu}, q^i, q^i_{\mu} \big){v}^{i} + \frac{\partial L}{\partial q^i _{\mu}}\big(x^{\mu}, q^i, q^i_{\mu}\big){v}^i_{\mu},
\end{equation*}
and the corresponding Euler-Lagrange field equations are: 
\begin{eqnarray}\label{eq: extendedfield-1}
\frac{d}{d x^{\mu}}\left(\frac{\partial \tilde{L}}{\partial q^i_{\mu}}\right) - \frac{\partial \tilde{L}}{\partial q^i}=0,  \qquad 
\frac{d}{d x^{\mu}}\left(\frac{\partial \tilde{L}}{\partial {v}^i_{\mu}}\right) - \frac{\partial \tilde{L}}{\partial {v}^i}=0,
\end{eqnarray}
or 
\begin{eqnarray}\label{eq: extendedfield-2}
\frac{d }{d x^{\mu}}\left[\frac{\partial^2 L}{\partial q^j\partial q^i_{\mu}}{v}^j
+\frac{\partial^2 L}{\partial q^j_{\gamma}\partial q^i_{\mu}}{v}^j_{\gamma}
\right] 
-\frac{\partial^2 L}{\partial q^i\partial q^j}{v}^j-
\frac{\partial^2 L}{\partial q^i\partial q^j_{\gamma}}{v}^j_{\gamma}&=&0\\
\frac{d}{d x^{\mu}}\left(\frac{\partial L}{\partial q^i_{\mu}}\right) - \frac{\partial L}{\partial q^i}&=&0
\end{eqnarray}
which corresponds to the Jacobi equations for $L$ (see \cite{CFT}). 

If we assume that $L$ is regular then $\tilde{L}$ is also regular since
\begin{equation}
  \det \begin{pmatrix}
  \frac{{\partial}^2 \widetilde{L}}{\partial q^i _{\mu}\partial q^j _{\gamma}}& \frac{{\partial}^2 \widetilde{L}}{\partial q^i _{\mu}\partial v^j _{\gamma}}\\
  \frac{{\partial}^2 \widetilde{L}}{\partial v^i _{\mu}\partial q^j _{\gamma}}&\frac{{\partial}^2 \widetilde{L}}{\partial v^i _{\mu}\partial v^j_{\gamma}}\end{pmatrix}=
  \det \begin{pmatrix}
  \frac{{\partial}^2 \widetilde{L}}{\partial q^i _{\mu}\partial q^j _{\gamma}}& \frac{{\partial}^2 {L}}{\partial q^i _{\mu}\partial q^j _{\gamma}}\\
  \frac{{\partial}^2 {L}}{\partial q^i _{\gamma}\partial q^j _{\mu}}&0\end{pmatrix}
  \not= 0
\end{equation}

Therefore, geometrically the Jacobi equation for the Lagrangian $ L: {\mathbb R}^k\times T^1_k Q\rightarrow {\mathbb R}$  can be written as following
\begin{align}\label{eq: Jacobi-geometric}
       dx^{\mu}(\Gamma_{\beta})&=\delta^{\mu} _{\beta}\; ,\qquad       1\leq \mu, \beta \leq k\\ 
       \sum_{\mu=1}^k i_{\Gamma_\mu} \Omega^{\mu} _{\tilde L}&=dE_{\tilde L} + \sum_{\mu=1}^k \frac{\partial \tilde{L}}{\partial x^{\mu}} dx^{\mu}    
      \end{align}
where $\Gamma\in \mathfrak{X}^k({\mathbb R}^k\times T^1_kTQ)$ and we obtain that the function $E_{\tilde L}$ is locally given by
\begin{align*}
E_{\tilde L}&=q^i_{\mu}\frac{\partial \tilde{L}}{\partial q^i_{\mu}}+v^i_{\mu}\frac{\partial \tilde{L}}{\partial v^i_{\mu}}-\tilde{L}\\
&=\frac{\partial^2 L}{\partial q^i_{\mu}\partial q^{j}}{v}^{j}q^i_{\mu} + \frac{\partial^2 L}{\partial q^i_{\mu}\partial q^j_{\gamma}}{v}^j_{\gamma}q^i_{\mu}-\frac{\partial L}{\partial q^i}v^i
\end{align*}
\section{A geometric variational description of dissipative field theories}

A modification of our technique for deriving the Jacobi equation allows us to give a variational description of field equations with dissipation described by the two data $(L, F)$ where $L: {\mathbb R}^k\times T^1_kQ\rightarrow {\mathbb R} $ and $F: {\mathbb R}^k\times T^1_k Q\rightarrow T^*(T^1_k)Q$ satisfy $\hbox{pr}_2=\pi_{(T^1_k)^*Q}\circ F$ where $\hbox{pr}_2: {\mathbb R}^k\times T^1_k Q\rightarrow T^1_kQ$ is the  projection onto the second factor. 
The term $F$ takes into account terms which in the field equations are not coming from a variational principle (such as dissipation or other external forces). 
Locally
\[
F=F_i({\bf x}, q^i, q^i_{\mu})\, dq^i +F^{\mu}_i({\bf x}, q^i, q^i_{\mu}) dq_{\mu}^i
\]

the field equations that we want to describe are
\begin{equation}\label{eq:forced}
\sum_{\mu}\frac{d}{d x^{\mu}}\left(\frac{\partial L}{\partial q^i_{\mu}}-F^{\mu}_i\right) - \frac{\partial L}{\partial q^i}=F_i. 
\end{equation}
\begin{proposition}
For the   Lagrangian 
$\tilde{L}_F: {\mathbb R}^k\times T^1_kTQ\rightarrow {\mathbb R}$ defined by
\begin{equation} 
\widetilde{L}_F\big(x^{\mu}, q^i, \dot{q}^{i}, v^i_{\mu}, \dot{v}^{i}_{\mu}\big)= \widetilde{L}\big(x^{\mu}, q^i, v^{i}, q^i_{\mu}, {v}^{i}_{\mu}\big)- F_j(x^{\mu}, q^i, q^i_{\mu})v^j-F^\gamma_j(x^{\mu}, q^i, q^i_{\mu})v^j_{\gamma}.
\end{equation} 
the corresponding Euler-Lagrange field equations are
\begin{eqnarray}\label{eq: extendedfield-1}
\frac{d}{d x^{\mu}}\left(\frac{\partial \tilde{L}_F}{\partial q^i_{\mu}}\right) - \frac{\partial \tilde{L}_F}{\partial q^i}=0,  \qquad 
\frac{d}{d x^{\mu}}\left(\frac{\partial \tilde{L}_F}{\partial {v}^i_{\mu}}\right) - \frac{\partial \tilde{L}_F}{\partial {v}^i}=0,
\end{eqnarray}
and
 the last equation is given exactly by equation (\ref{eq:forced}). 
 \end{proposition}

\begin{example}
In \cite{Baggioli} the authors developed a field theory with dissipation and analyzed the properties of the Lagrangian and the Hamiltonian with two fields. Following we give an example which the Lagrangian and  dissipation  are used there.

 Consider a Lagrangian $L\in C^{2}\big( \r^2 \times T_2 ^1 Q \big)$ 
 \begin{equation*}
  L(t,x; q, q_t, q_x)=\frac{1}{2}\big(q_t ^2-c^2 q_x^2\big)   
\end{equation*}
with  corresponding  Euler-Lagrange field equations 
\begin{equation*}
\frac{d}{d t}\left(\frac{\partial L}{\partial q_t}\right) + \frac{d}{d x}\left(\frac{\partial L}{\partial q_x}\right) - \frac{\partial L}{\partial q}=0
\end{equation*}
which leads to
\begin{equation}\label{9}
\frac{d}{d t}(q_t) + \frac{d}{dx}(-c^2q_x)\equiv  q_{tt}- c^2q_{xx}=0.
\end{equation}
Now, by taking into account the Maxwell interpolation, the previous equations may also include a dissipative term: 
\begin{equation*}
   c^2 \frac{\partial^2 q}{\partial x^2}=\frac{\partial^2 q}{\partial t^2}+\frac{1}{\tau}\frac{\partial q}{\partial t},
\end{equation*}
To obtain a pure variational formulation we consider the Lagrangian: 
\begin{align*}
 \tilde{L}_F(x, y, q, v, q_t, q_x, {v}_t, {v}_x)= 
 \frac{\partial L}{\partial q^{i}}\big(x^{\mu}, q^i, q^i_{\mu} \big){v}^{i} + \frac{\partial L}{\partial v^i _{\mu}}\big(x^{\mu}, q^i, q^i_{\mu}\big){v}^i_{\mu}
 -Fv, 
\end{align*}
 is obtained. In this case the dissipation term is
\begin{equation*}
F: \r^2 \times (TQ \oplus TQ)\rightarrow T^{\star} Q, \ \ \ \ 
 F(t, X, q, q_t, q_x)=-\frac{1}{\tau}q_t.
\end{equation*}
and 
\begin{equation}
\tilde{L}_F=
q_t{v}_t- c^2q_x{v}_x + \frac{1}{\tau} q_t v.
\end{equation}
Therefore  the Euler Lagrange field equations for  $\widetilde{L}_F$ are
\begin{equation}
 \begin{array}{cc}
     q_{tt}-c^2q_{xx}-\frac{1}{\tau}q_t=0 &\\ \\
     {v}_{tt}-c^2{v}_{xx}+\frac{1}{\tau}v_t=0. &
 \end{array}
 \end{equation}
\end{example}
In fact, this study provides an alternative way of deriving the energy density formula, where they have never been derived from the more systematic approach of the Lagrangian-Hamiltonian scheme. Furthermore, the author \cite{pluan} introduced the Lagrangian density and the dissipation function density for the Lagrangian description of electrodynamics. It was shown that the Hamiltonian densities corresponding to this are identical to the energy densities derived in the Hamiltonian scheme.  
\subsection{Conclusions and Future work}
In this paper we have introduced a geometric framework for the Jacobi equation for field theories and a variational extension for field theories with dissipation term. As a future work, it is possible to develop similar ideas in a multisymplectic framework that is more general that the one presented in this paper. The extension to Hamiltonian field equations with dissipation and reduced theories \cite{Leo-mar-vi-sa,martin-vila} it is an interesting topic of future study. Finally, applications to the derivation of multisymplectic integrators for field theories with dissipation using well established  techniques for the purely variational case \cite{marsden-skoller}.

\

{\bf Acknowledgments:}  DMdD  acknowledges the financial support from the Spanish Ministry of Science and Innovation under grants PID2022-137909NB-C21, PCI2024-155047-2 and  from the Severo Ochoa Programme for Centres of Excellence in R\&D (CEX2023-001347-S). 

\bibliography{References}
\end{document}